\documentclass[useAMS,usenatbib]{mn2e}

\usepackage{graphicx} 
\usepackage{pslatex}
\usepackage{bm}

\title[Environment influence on strong lenses]{Background, foreground
and nearby matter influence on strong gravitational lenses}
\author[M. Jaroszynski and Z. Kostrzewa-Rutkowska]
{M. Jaroszynski$^{1}$\thanks{E-mail:
mj@astrouw.edu.pl (MJ); zkostrze@astrouw.edu.pl (ZKR)} 
and Z. Kostrzewa-Rutkowska$^{1}$\\
$^{1}$University of Warsaw Observatory, Al. Ujazdowskie 4, 00-478 Warsaw,
Poland}

\begin{document}

\date{Accepted 2012 April 26; Received 2012 April 25; in original form 
2011 November 10}

\pagerange{\pageref{firstpage}--\pageref{lastpage}} \pubyear{2012}

\maketitle

\label{firstpage}

\begin{abstract}
We investigate strong lensing by non-singular finite isothermal
ellipsoids taking into account the influence of the matter along the
line of sight and in the close lens vicinity. We compare three descriptions
of light propagation: the full approach taking into account all matter
inhomogeneities along the rays, the single plane approach, where we take into
account the influence of the strong lens neighbours but neglect the
foreground and background objects, and the single lens approach. In each
case we simulate many strong lensing configurations placing a point
source at the same redshift but in different locations inside the region
surrounded by caustics. We further analyze configurations of four or
five images. For every simulated strong lensing configuration we attempt 
to fit a simplified lens model using a single isothermal ellipsoid 
or a single isothermal ellipsoid with external shear. The single lens
fits to configurations obtained in the full approach are rejected in 
majority of cases with 95\% significance. For configurations obtained 
in the single plane approach the rejection rate is substantially lower.
Also the inclusion of external shear in simplified modeling improves 
the chances of obtaining acceptable fits, but the problem is not solved 
completely. The quantitative estimates of the rates of rejection of 
simplified models depend on the required accuracy of the models, and 
we present few illustrative examples, which show that both matter close
to the lens and matter along the rays do have important influence on 
lens modeling. We also estimate the typical value of the external 
shear and compare the fitted parameters of the simplified models 
with the parameters of the lenses used in the simulations.
\end{abstract}


\section{Introduction}

Strong gravitational lensing of distant sources has low probability
and is usually attributed to a single concentration of matter close to
the line of sight, while the influence of anything else is neglected
(see e.g. \citealp{K06} for a review of strong lensing). In some cases
simple, single component models of lenses are not satisfactory. 
Even the first lensed quasar observed, 
QSO 0957+561 A,B \citep{b32,b33}  requires
a combination of a galaxy and a galaxy cluster to model its basic
properties \citep{Y81}.

The influence of another object along the line of sight on the main lens
properties has been investigated by \citet{KA88}. \citet{KKS97} argue
that adding an external shear to an elliptic lens model greatly improves
the fits of four image configurations, but they conclude that the
shear is probably related to the main lens. \citet{BK96} investigates
the influence of the density fluctuations along the line of sight on the
accuracy of Hubble constant determination based on time delays measured
in some of the multi-image QSO systems. \citet{ChKK03} check the
influence of the substructure not related to the main lens on the
observed flux ratios in lensed images. \citet{WBO04} study the
probability of giant arcs using the results of a large cosmological
simulation and multi-layer approach to backward ray tracing. They find
that the matter on the way slightly increases the probability of arcs.
\citet{WBO05} address specifically the question of secondary lenses on
the line of sight. According to this study the role of secondary lenses
is a strong function of source redshift: it can be neglected in 95\% of
cases for a source at $z_\mathrm{S}=1$, 
but is important in 38\% of cases at
$z_\mathrm{S}=3.8$. Using single plane approach introduces
systematic errors to the estimated lens parameters. In a recent paper
\citet{AN11} investigate the influence of the secondary lenses in the
line of sight on the accuracy of estimating the dark energy density.

It is probably impossible to quantify the influence of the matter along
the line of sight and of the strong lens close (but possibly
unobservable) neighbours in a single case of real observed multiple
image configuration. In this paper we are going to obtain a statistical
measure of such an influence performing simulations of many multiple
image configurations using a realistic model of light propagation in an
inhomogeneous Universe model. To find the role of matter not belonging to
the strong lens, we follow the rays including or not the lens neighbours
and inhomogeneities along the rays. Again, the direct comparison between
individual image configurations obtained as a result of different
descriptions of light propagation seems useless, since one should
compare positions and flux ratios of images of a point source {\it at
the same position} and it is not clear how to compare source positions
between the different models of light propagation. If image
configuration is considered, it is the source position relative to the
caustic structure which counts and since the caustic structures are
different the problem persists. 

Instead of making comparisons on a one to one basis, we compare the
properties of all four or five image configurations between the models
of light propagation. If the influence of matter outside the strong lens
were unimportant, each lens - image configuration would be similar to
some configuration obtained with a single lens model. To check whether
this is the case we attempt to fit such simplified model to all image
configurations considered. The success rate of the fitting procedure
applied to configurations obtained with a different model of light
propagation statistically measures the similarity of this model to
the single lens model. 

In Sec.~2 we describe our models of light propagation. Sec.~3  presents
tools used to compare different models and the results of such
comparison. Discussion and conclusions follow in Sec.~4.

\section[]{Model of the light propagation}

\subsection{Deflection angles}

The results of the Millennium Simulation \citep{s05}
give the matter distribution 
(the positions and velocities of $\sim 10^{10}$ simulation particles)
in the 64 epochs corresponding to the redshifts $0 \le z_i \le127$. 
In our calculations we assume, that the propagation of a ray between
redshifts $z_{i_1} \equiv (z_{i-1}+z_i)/2$ and 
$z_{i_2} \equiv (z_{i+1}+z_i)/2$ is affected by matter 
distribution given for the epoch $z_i$. We are interested in sources at high
redshifts ($z \ge 1$), so there are always many such matter layers on the 
ray path, and each of them can be  considered thin. Thus the multilayer 
approach \citep{b1} with layers corresponding to the 
Millennium epochs is natural. The position of the ray in the $N$-th 
layer in the angular coordinates is given as
\begin{equation}
\bm{\beta}_N = \bm{\beta}_1 
- \sum_{i=1}^{N-1}~\frac{d_{iN}}{d_{N}}~\bm{\alpha}_i(\bm{\beta}_i)
\end{equation}
where $d_{iN}$ is the angular diameter distance as measured by an observer
at epoch $i$ to the source at epoch $N$, $d_N$ - the angular diameter
distance to the same source measured by a present ($z=0$) observer,
and $\bm{ \alpha}_i(\bm{ \beta}_i)$ is the deflection angle in the 
$i$-th layer at the position $\bm{\beta}_i$. In a flat cosmological model
the angular diameter distances in the lens equation can be replaced by 
comoving distances. Below we use comoving distances $D(z)$, denoting 
$D_i \equiv D(z_i)$ for short.  Since the comoving distances are 
additive in a flat model, one has $d_{iN}/d_N=(D_N-D_i)/D_N$. In the
calculations we apply more efficient recurrent formula of \citet{b34}, 
equivalent to the above equation.

To calculate the deflection angle in a given layer we need a description
of  its matter distribution. This is done in two steps.
The averaged matter density is defined on a coarse grid of $256^3$ cells. 
The gravitationally bound haloes have been described by  \citet{b11}
and  \citet{b4}. Both kinds of data are 
accessible from the Virgo - Millennium Database \citep{b23}.
\citet{JK10} use the same information on matter distribution, but their
calculation of lensing effect is based on different approach.

Since only the positions, virial masses, virial radii, and virial 
velocities of the haloes can be obtained from the Database, the detailed matter
distributions of each halo must be supplemented. 
We use non singular isothermal 
ellipsoids (NIE) as models of individual lenses. 
The averaged matter density 
includes also the matter belonging to haloes, so the halo mass models must be 
compensated by a shallow negative density distribution (see below).

In constructing the matter layers we follow \citet{b10}, randomly 
rotating and shifting simulation cubes corresponding to different epochs, 
which eliminates the consequences of periodic boundary conditions used in 
their calculations. The comoving thickness of the $i$-th layer 
$D_{i_2}-D_{i_1}$ is of the order of $10^2~\mathrm{Mpc}$, 
always smaller than the simulation cube size $500/h~\mathrm{Mpc}$.
The 2D layers are not periodic in general. 
\citet{b17} define specific projection directions to obtain periodic 2D 
density distributions and use 2D Fourier transforms in calculations 
of the deflection angles. Our 3D grid ($256^3$) is small enough to employ 
spectral methods. Using Poisson equation we calculate 3D gravitational 
acceleration ${\bm g}$ on the grid. The deflection of a ray passing the 
$i$-th layer at position $\bm{\beta}_i$ due to the averaged (or {\it background})
matter distribution is given as:
\begin{equation}
\bm{\alpha}_i^\mathrm{bcg}(\bm{\beta}_i) = \frac{2}{(1+z_i)c^2}~
\int_{D_{i_1}}^{D_{i_2}}~{\bm g}_\perp(D,\bm{\beta}_i)~dD
\end{equation}
where ${\bm g}_\perp$ is the component of gravitational acceleration 
perpendicular to the ray, and its value at any location is obtained by the 
interpolation on the 3D grid.
$(D,\bm{\beta})$ are used as coordinates. Since our simulated maps 
of the sky cover typically regions of a few minutes of arc in size, the 
position components $(D\beta_x,D\beta_y)$ may be treated as Cartesian 
coordinates
in the plane perpendicular to the propagation direction, and the comoving 
distance $D$ serves as radial coordinate. Because the integration should be 
performed over the proper distance, instead of comoving one, the result 
is divided by the factor $1+z_i$.

For each individual halo we apply the model of non-singular isothermal
ellipsoid as described by \citet{K94}. The deflections
in 2D real notation are given by \citet{K06}:
\begin{equation}
\alpha_x=
\frac{\alpha_0}{q^\prime}~\mathrm{arctan}
\left(\frac{q^\prime~x}{\omega+r_0}\right)~~
\alpha_y
=\frac{\alpha_0}{q^\prime}~\mathrm{artanh}
\left(\frac{q^\prime~y}{\omega+q^2r_0}\right)
\end{equation}
where:
$$
\omega(x,y,q,r_0)=\sqrt{q^2(x^2+r_0^2)+y^2}~~~~q^\prime=\sqrt{1-q^2}
$$
The ray crosses the lens plane at $(x,y)$, the lens centre is placed at
the  origin of the coordinate system, the major axis along $x$. 
The axis ratio is given by $q$, $r_0$ is the core radius, and $\alpha_0$
is  the deflection angle parameter (see below).

The convergence $\kappa$ (which is proportional to the surface mass
density) has the required (NIE) form:
\begin{equation}
\kappa(x,y,q,r_0)=\frac{1}{2}(\alpha_{x,x}+\alpha_{y,y})
=\frac{1}{2}~\frac{\alpha_0}{\omega(x,y,q,r_0)}
\end{equation}
Far from the centre the 2D matter distribution becomes elliptical, as
required. 

The NIE matter distribution is infinite with divergent total mass. To
obtain a finite mass lens we follow \citet{K06} combining two NIE
distributions  with different core radii $r_1 \ll r_2$. Inside $r_1$ the
surface density is approximately constant, isothermal in between, and
sharply falling outside $r_2$:
\begin{equation}
\kappa(x,y,q)=\kappa(x,y,q,r_1)-\kappa(x,y,q,r_2)
\rightarrow \frac{1}{4}~\frac{\alpha_0(r_2^2-r_1^2)}{\omega^3(x,y,q,0)}
\end{equation}
which gives a non-singular, finite mass lens with the deflection angle:
\begin{equation}
\bm{\alpha}(x,y,q)=\bm{\alpha}(x,y,q,r_1)-\bm{\alpha}(x,y,q,r_2)
\rightarrow \alpha_0(r_2-r_1)\frac{{\bm r}}{r^2}
\end{equation}
For large radii $r \gg r_2$ the lens acts as a spherical finite mass 
contained within $r_2$.
We identify the mass and radius  of the lens with  the virial mass and
radius  of the halo, so the characteristic deflection angle is given as:
\begin{equation}
\alpha_0=
\frac{4GM_\mathrm{vir}}{c^2 (r_2-r_1)} \approx
\frac{4GM_\mathrm{vir}}{c^2 r_\mathrm{vir}}
~~~~~~~~r_2=r_\mathrm{vir}
\end{equation}

To compensate the lens we use
a circular disk of radius $r_\mathrm{lim}\gg r_\mathrm{vir}$ with constant 
surface density and with the deflection angle:
\begin{equation}
\bm{\alpha}_\mathrm{cmp}({\bm r})
=\frac{4GM_\mathrm{vir}}{c^2}\times\left\{ 
\begin{array}{ll}
   {\bm r}/r_\mathrm{lim}^2 & \mbox{if $r\le r_\mathrm{lim}$}\\
   {\bm r}/r^2              & \mbox{if $r> r_\mathrm{lim}$}
\end{array}
\right.
\end{equation}
where
\begin{equation}
M_\mathrm{vir} \equiv \frac{4}{3}\pi\rho_0r_\mathrm{lim}^3
\end{equation}
defines the size of the disk. $\rho_0$ is the average matter density 
in the Universe at the lens redshift. In our approach we always use 
$r_1=0.01r_2$, so we avoid singularity at the lens centre, but we do 
not assign any physical meaning to this parameter.

Finally we check numerically the dependence of the deflection angle 
on various parameters. We use the formula for a compensated lens:
\begin{equation}
\bm{\alpha}({\bm r})=\bm{\alpha}(x,y,q,r_1)
   -\bm{\alpha}(x,y,q,r_2)-\bm{\alpha}_\mathrm{cmp}({\bm r})
\end{equation}
\begin{figure}
\includegraphics[width=84mm]{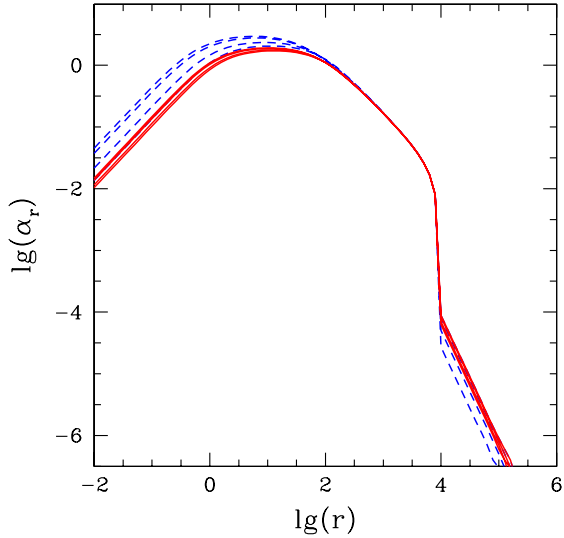}
\caption{The deflection angle of the compensated lens. For illustration we use:
$r_1=1$, $r_2=10^2r_1$, and $r_{lim}=10^2r_2$. 
The plots show the radial deflection component $\alpha_r$
as function of radius for fixed polar angle $\phi=90^\circ$, $60^\circ$,
$30^\circ$, and $0^\circ$ (top - down).
Dashed  lines are used for $q=0.3$, and solid lines for $q=0.7$.}
\label{alfa-r}
\end{figure}
We illustrate the dependence of deflection angle on position in 
Fig.~\ref{alfa-r}, using arbitrary units and a case with 
$r_1 \ll r_2 \ll  r_\mathrm{lim}$. The compensation is not exact, 
but at $r\approx r_\mathrm{lim}$ the deflection angle drops by 
two orders of magnitude.

A compensated lens has only a finite range. Its radius depends on the
properties of each individual halo, but investigating the deflection by
haloes at any position in a given layer we take into account only the
finite number of lenses:
\begin{equation}
\bm{\alpha}^\mathrm{haloes}_i(\bm{\beta})
=\sum_j~\bm{\alpha}(\bm{\beta}-\bm{\beta}_j)
~~\mathrm{where:}~~D_i|\bm{\beta}-\bm{\beta}_j| \le r^\mathrm{lim}_j
\end{equation}
The expression $\bm{\alpha}(\bm{\beta}-\bm{\beta}_j)$ stands for 
the deflection at the position $\bm{\beta}$ 
caused by the lens at position $\bm{\beta}_j$
within its limiting range.

The total deflection in the $i$-th layer
\begin{equation}
\bm{\alpha}_i(\bm{\beta}_i) = 
   \bm{\alpha}^\mathrm{bcg}_i(\bm{\beta}_i) 
  +\bm{\alpha}^\mathrm{haloes}_i(\bm{\beta}_i) 
\end{equation}
can be calculated for a ray passing at any location $\bm{\beta}_i$.

\subsection{Maps of the sky}

We calculate  deflection angles for all layers of interest
storing them on the $2048^2$ grids with $0.1~\mathrm{arcsec}$
resolution. Using a set of stored deflection angles for all layers 
and interpolating on the grids one can apply Eq.(1), mapping a region of
$\approx 3^\prime \times3^\prime$ in the sky. We investigate light
propagation from sources at $z_S\approx 2$ taking into account the
influence of  several thousands haloes in the field of view and some in 
adjacent regions, if they are within their $r_\mathrm{lim}$ ranges.
We construct eight separate maps pointing in random directions.

When investigating multiple image properties we need zoomed maps of
smaller parts of the sky. For better resolution we use finer grids with
deflections interpolated from previous calculations with the help of
bi-cubic spline, so the interpolated deflection derivatives are
continuous. The backward ray shooting on a finer grid is repeated.

Treating each of the individual haloes in the field of view as an isolated
lens of known velocity dispersion and using SIS model we calculate its
Einstein radius $r_E$ for a source at given redshift $z_S$ in a homogeneous
Universe model. In each simulated {\it map} of the sky we find 10 lenses 
with the largest Einstein radii. 
We expect that these {\it dominating} lenses have the highest probability 
of producing  multiple images.

In the single SIS lens case the second image is possible only for a source 
within the Einstein ring. The brighter of the two images lies within $2~r_E$
from the lens centre and the dimmer within $1~r_E$,  so we expect that
in more complicated problem, including lens ellipticity and the influence 
of other lenses the multiple images lie within few Einstein 
radii from each other. 

We check for the presence of other haloes inside  the circle of the radius 
$3~r_E$ surrounding a dominating lens. If they are present we enlarge the 
region of interest including $3~r_E$ zones around all companions. Finally 
we repeat backward ray shooting inside a square on the sky including the 
region of interest. The fine grids giving the deflection angles in consecutive
layers encompass still larger areas, so one can follow rays deflected off
the main region. The result of the ray shooting is a vector array:
\begin{equation}
\bm{\beta}_N^{kl}=\bm{\beta}_N(\bm{\beta}_1^{kl})
\end{equation}
where $\bm{\beta}_N^{kl}$ gives the positions in the source plane 
of rays apparently coming from the directions $\bm{\beta}_1^{kl}$ 
on the observer's sky. Superscripts $k$, $l$ enumerate the rays. 
The $\bm{\beta}_N(\bm{\beta}_1)$ relation is given in Eq.(1).

\subsection{Three propagation models}

\begin{figure*}
\includegraphics[width=176mm]{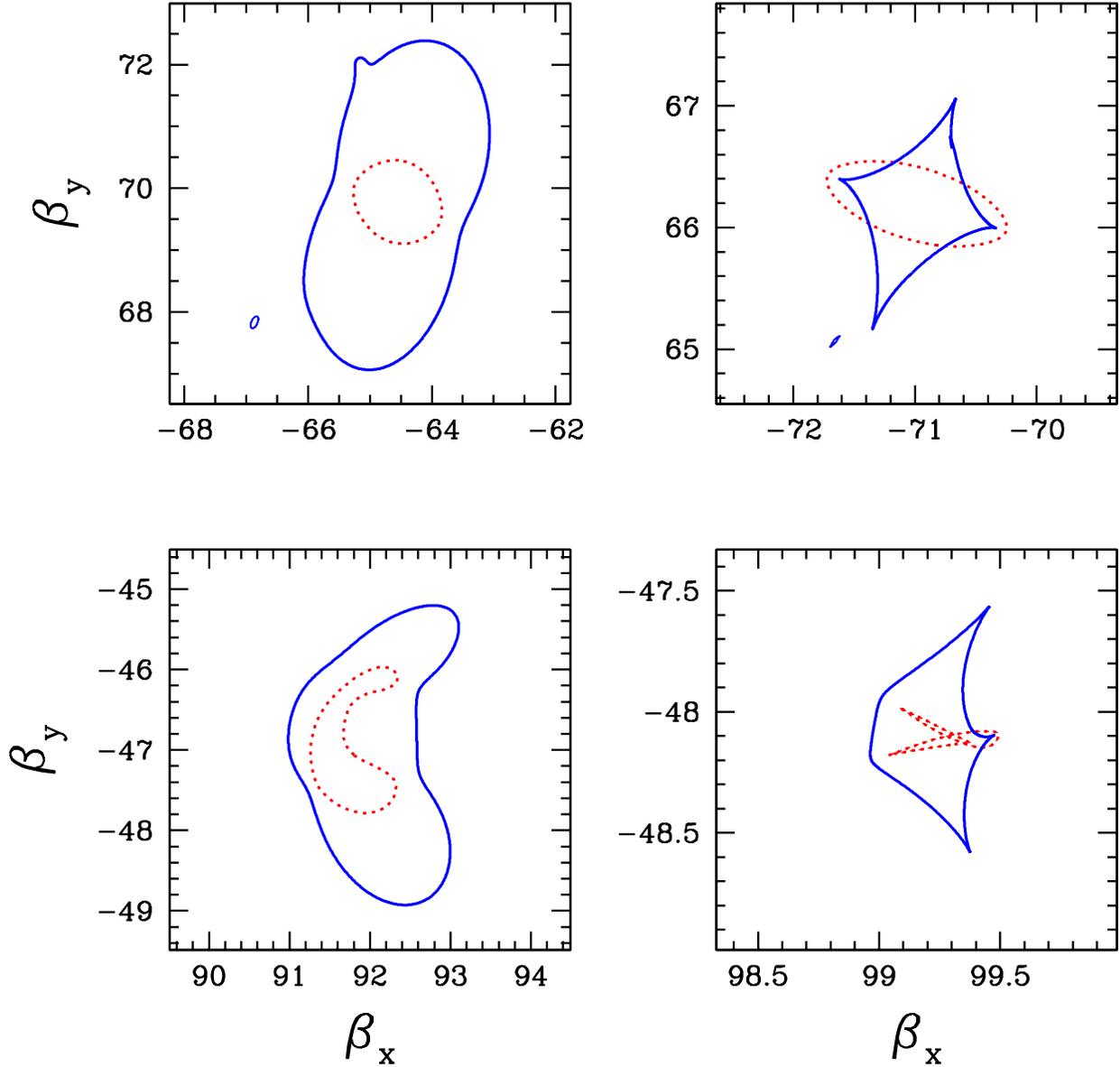}
\caption{Two examples of critical lines and caustic structures obtained 
using {\it Case 1} light propagation model. Critical lines are shown 
in the left column.  The coordinates are relative to the centre of 
$3^\prime\times 3^\prime$ maps and are expressed in arcseconds. The 
corresponding caustic structures are shown on the right. The perturbations 
of the caustic structure by the surrounding haloes are apparent in the 
example shown in the upper row. In the lower row the action of two 
close elliptical lenses of similar masses, having different position 
angles completely changes the shapes of critical lines and caustics.}
\label{panel}
\end{figure*}

To estimate the influence of the foreground and background matter in the 
beam and also of the lens neighbours belonging to the same redshift layer
we consider three scenarios for the backward ray shooting. Very similar
method is employed by \citet{b26}. 
In the {\it full} approach 
({\it Case 1}) we include deflection of rays by matter in all layers 
between the source and observer, as described above. In the {\it one layer} 
approach ({\it Case 2}) we include deflection by all matter in the layer 
of dominating lens only, neglecting the influence of other layers. 
In the  {\it single lens} approach ({\it Case 3}) only the deflection 
by the dominant lens is taken into account. 

\subsection{Image finding}

To find critical lines in the sky we solve the standard equation 
\citep{SEF92}:
\begin{equation}
\det~\mathsf{A}=0~~~~~~~~~~~~
\mathsf{A}\equiv\left|\left|\frac{\partial\bm{\beta}_N}
      {\partial\bm{\beta}_1}\right|\right|
\end{equation}
where $\mathsf{A}$ is the deformation matrix. The equation is solved 
approximately on the grid and then refined by iterations. 
Finding of caustic lines in the source plane is then 
straightforward.

Usually the critical lines and caustic structures obtained in
three {\it Cases} of backward ray shooting are similar and the 
influence of matter outside the dominating lens is not apparent. There
are however several cases showing the influence of smaller mass haloes on the 
critical lines and (to a lesser extent) caustics as shown in the upper row
of Fig.~\ref{panel}. An example of a strong caustic perturbation is shown 
in the lower row of Fig.~\ref{panel}.

We start image finding using approximate methods on the grid. We replace 
the point source by an extended surface brightness profile with the Gaussian 
shape and the characteristic radius of few pixels. The related surface 
luminosity  in the sky is given by:
\begin{equation}
I^\mathrm{obs}(\bm{\beta}_1)=I^\mathrm{src}(\bm{\beta}_N(\bm{\beta}_1))
\end{equation}
The local maxima of the observed surface luminosity are the positions of the 
images of the source centre. We find brightness maxima on the grid and use
them as approximate solutions to the lens equation. The improved positions 
are obtained by iterations.

Using the above method we find multiple images of point sources placed 
behind lenses. For each lens we try few hundreds to several thousands source
positions (depending on the size of the caustics region) distributing
them evenly. Each position considered represents the same
surface area in the source plane. Finding a source at any location has 
the same probability if they are distributed randomly. (We neglect here
the amplification bias caused by the strong lensing.)
The procedure of image finding for many source positions is repeated for 
each of the ten strongest lenses belonging to each of the eight simulated 
maps of the small fragments of the sky obtained with the three different 
models of light propagation.

\section{Fitting the simulated strong lenses with simplified models}

In the case of pure elliptical, 
non-singular lens one would expect (in general) 1, 3, or 5 images. Our 
numerical method using at the beginning a small extended source 
and having finite resolution may miss some of the images. (See the 
Discussion). In {\it Cases 1,2} 
the external influence may also change the number of images. 
Since configurations with more images probably better constrain the lens, 
we concentrate on cases with five or four images.

To estimate the influence of the galaxies in the lens vicinity and the 
matter along the line of sight on the properties of the strong lensing, 
we attempt to model all our simulated cases of multiple imaging using a 
single lens model in a uniform Universe. The single lens we are using 
in modeling is a non-singular finite isothermal ellipsoid used also in the 
simulations for each of the haloes. When tracing the rays in simulations we 
interpolate all deflection angles from earlier computed arrays. Single 
lens modeling uses analytically calculated deflection angles and their 
derivatives, so it serves also as a test of ray tracing simulations in 
{\it Case 3}.

We also try a more sophisticated lens model using the same non-singular
finite isothermal ellipsoids with external shear. In this way the tidal
influence of masses close to the rays is at least partially represented. 

Simulations give the lens position, image positions and image 
amplifications corresponding to their energy fluxes.
These parameters would also be observed in reality. 
The source position $\bm{\beta}_\mathrm{S}$ is unknown
but it must be the same for all the images, and we use it as 
a model parameter. The source luminosity is also unknown, but the
observed fluxes must be proportional to the lens amplifications in
corresponding positions. Thus the model has to reproduce amplification
ratios and not their absolute values.
The intrinsic lens parameters (axis ratio $q$,
characteristic deflection angle $\alpha_0$, the virial radius 
$r_\mathrm{vir}$, 
and the lens position angle in the sky $\phi$) are not so easy to measure 
and we treat them as unknown model parameters. Also the shear components
$\gamma_1$, $\gamma_2$ belong to the class of intrinsic parameters. Below
we use the value of shear defined as
\begin{equation}
\gamma \equiv \sqrt{\gamma_1^2+\gamma_2^2}
\end{equation}

We treat the lens position, image positions and image fluxes ratios given 
by the simulations as observed quantities, each with its own uncertainty.
We attempt to fit every simulated strong lensing case with our simplified 
model. We are going to reproduce the observed lens position, 
image positions and image flux ratios (or differences in image 
stellar magnitudes) looking for a minimum of $\chi^2$ taking all three 
constraints into account:
\begin{equation}
\chi^2=\chi^2_\mathrm{L}+\chi^2_\mathrm{I}+\chi^2_\mathrm{m}
\end{equation}
The first term, controlling the lens position, used in the model has 
the obvious form:
\begin{equation}
\chi^2_\mathrm{L}=
\frac{(\bm{\beta}_\mathrm{L}-\bm{\beta}_\mathrm{L}^{~0})^2}
{\sigma_\mathrm{L}^2}
\end{equation}
where the subscript $\mathrm{L}$ stands for ``lens'' and 
$\sigma_\mathrm{L}$  is the assumed accuracy of the lens position.

Using the simplified model we can calculate the source positions 
$\bm{\beta}_\mathrm{S}^{~i}$ related 
to observed image positions 
$\bm{\beta}_\mathrm{I}^{~i0}$.
The deformation matrix $\mathsf{A}_{(i)}$ 
and magnification matrix $\mathsf{M}_{(i)}$ can be calculated at each 
image position:
\begin{equation}
\bm{\beta}_\mathrm{S}^i
\equiv\bm{\beta}_\mathrm{S}(\bm{\beta}_\mathrm{I}^{~i0}) 
~~~~~~~~
\mathsf{A}_{(i)}\equiv\left|\left|\frac{\partial\bm{\beta}_\mathrm{S}^{~i}}
{\partial\bm{\beta}_\mathrm{I}^{~i0}}\right|\right|
~~~~~~\mathsf{M}_{(i)}\equiv\mathsf{A}_{(i)}^{-1}
\end{equation} 
The mismatch between $\bm{\beta}_\mathrm{S}^{i}$ and 
$\bm{\beta}_\mathrm{S}$ 
implies magnified mismatch between modeled and observed image positions, 
which gives for $\chi^2_\mathrm{I}$ (\citealp{K06}):
\begin{equation}
\chi^2_\mathrm{I}=\sum_i~
\frac{\left|\mathsf{M}_{(i)}\mathbf{\cdot}
~\left(\bm{\beta}_\mathrm{S}^i-\bm{\beta}_\mathrm{S}\right)\right|^2}
{\sigma_\mathrm{I}^2}
\end{equation}
The fitting statistic for flux ratios is given as:
\begin{equation}
\chi^2_\mathrm{m}=\sum_i
\frac{\left(m^i-m^{0i}-\left<m^i-m^{0i}\right>\right)^2}
{\sigma_\mathrm{m}^2}
\end{equation}
In all cases the quantities with extra superscript ``$0$'' are taken 
from simulations and mimic the observed values while 
the quantities without this superscript are given 
by the model. 

We use the following characteristic values for the accuracy parameters: 
$\sigma_\mathrm{I}=0.01~\mathrm{arcsec}$,
$\sigma_\mathrm{L}=0.1~\mathrm{arcsec}$,
and $\sigma_\mathrm{m}=0.1$ or $10$ (in stellar magnitudes)
This choice takes into account the fact that image positions are easier 
to measure  than the lens position. 
With the small value of the flux measurement 
error ($\sigma_\mathrm{m}=0.1$) we use the fluxes as important factor 
in modeling. The other value makes the fluxes unimportant in the model 
fitting. Using it one can check to which extent models well 
reproducing image positions are also good in reproducing ratios of their 
fluxes. 

We start fitting process using a model with all parameter values, also the 
intrinsic lens parameters,  taken from the simulation. This is not a general 
approach, but it is fast in finding solutions close to the starting point 
in parameter space (if they are present). 
It is also our goal to obtain the ``observable'' 
parameters as close to simulated values as possible, but the 
intrinsic lens parameters are not constrained a'priori and have no 
preferred values. 

Our approach to model fitting is not general, since we are probing only 
the small region of parameter space and there is a chance that we miss some 
acceptable models of the lens. On the other hand, if the influence of the 
environment were weak, the resulting lens model should be close to the 
original lens used in the simulations. Failure to find a model which 
is similar to the original can be interpreted as the result of the strong
influence of the matter in the rays vicinity.

The number of the degrees of freedom ($DOF$) depends on the number of
images ($N_\mathrm{im}$) and on the number of intrinsic model
parameters, which is $N_\mathrm{par}=4$ for an elliptic lens 
($q$, $\alpha_0$, $r_\mathrm{vir}$, $\phi$) and $N_\mathrm{par}=6$ 
if the external shear ($\gamma_1$, $\gamma_2$) is taken into account.
The lens position is both observed and modeled, so it has no impact.
The source position requires 2 parameters, and the source internal luminosity
(which we do not use explicitly in modeling) - another one.
When modeling the flux ratios with $\sigma_\mathrm{m}=0.1$ we effectively 
use the magnification information from the simulations, but for 
$\sigma_\mathrm{m}=10$ it is practically lost. 
In calculations we use: 
$DOF=3(N_\mathrm{im}-1)-N_\mathrm{par}$ (if $\sigma_\mathrm{m}=0.1$) or
$DOF=2(N_\mathrm{im}-1)-N_\mathrm{par}$ (if $\sigma_\mathrm{m}=10$).
(That implies that modeling of four-image configurations including shear 
and neglecting flux ratios is impossible, since $DOF=0$ in this case).
Using tables of $\chi^2$ distribution we reject fits if they give
\begin{equation}
\chi^2 > \chi_{0.95}^2(DOF)
\end{equation}
where $\chi_{0.95}^2$ corresponds to 95\% significance for a given $DOF$.

The influence of the matter distribution outside the lens on the lensing
process can be characterized  by the difficulty in obtaining an acceptable
single lens model in such more complicated situations.
To measure the rate of success in fitting procedure we assign the same
weight to each investigated source position since they are distributed
evenly in the source plane. Thus the more massive lenses have greater 
influence on the results as compared with less massive. This is the 
simplest consequence of the observational selection. More subtle effects, 
like  magnification bias, influence of image configuration etc are too 
difficult to model within our approach. 

Since we perform our simulations in eight different, randomly chosen 
regions of the synthetic Universe, we can estimate the role 
of the cosmic variance in our simulations.
Calculating the rate of success for each of the regions separately 
and finding the dispersion of the results, we get an error estimate.

\begin{table}
\centering
 \begin{minipage}{84mm}
  \caption{Acceptability of fits - dependence on the model}
  \begin{tabular}{@{}ccccc@{}}
  \hline
  Shear 	&  no &  no &  yes &  yes\\
 $\sigma_\mathrm{m}$ & 0.1 & 10.  & 0.1 & 10.\\
 \hline
 {\it Case 1} & $0.08\pm0.13$ & $0.11\pm0.13$ & $0.43\pm0.17$ & $0.68\pm0.22$\\
 {\it Case 2} & $0.38\pm0.21$ & $0.41\pm0.18$ & $0.59\pm0.22$ & $0.78\pm0.19$\\
 {\it Case 3} & $.9998\pm .0002$ &$.9998\pm.0001$& $.9997\pm.0002$ & $.9998\pm.0001$\\
 \hline
\noalign{\vskip3pt}
\multicolumn{5}{p{8.4cm}}{Note: The table shows the dependence of 
the rate of acceptability of fits on the method of treating the light
propagation ({\it Cases 1 -- 3}, see the text for details),  
for models neglecting (``no'') or taking into account (``yes'') 
the external shear and reproducing ($\sigma_\mathrm{m}=0.1$) or
not ($\sigma_\mathrm{m}=10$) the flux ratios.
The assumed errors in fitting the positions of the lens center and 
of the images are kept constant 
($\sigma_\mathrm{L}=0.1$,  $\sigma_\mathrm{I}=0.01$). }
\end{tabular}
\end{minipage}
\label{accept1}
\end{table}

In Table.~1 we show the chances that a simulated strong lensing 
case has an acceptable simplified model. In {\it Case 3} acceptable 
models are possible to obtain in almost 100\% of simulated lens-image 
configurations, which  shows that our fully numerical (used in obtaining 
lens - image configurations) and semi-analytical (used in fitting models) 
descriptions of the same physical approach are in complete agreement.

The results depend on the adopted values of position accuracy,  (see 
below) but the difficulty in modeling realistic configurations using 
a single elliptical lens without external shear is apparent, even if 
the reproduction of correct flux ratios is not required. 
The inclusion of external shear substantially improves the chances 
of obtaining acceptable solutions with simplified lens models, 
but still there is no guarantee of success, especially if one requires
the reproduction of the observed flux ratios.

The present day astrometry gives better accuracy of image positions than 
assumed in majority of our simulations. On the other hand the micro-lensing 
may change the measured fluxes of individual images in a random way, which
diminishes their role as model constraints. In Table.~2 we show the results
of our approach with increased required accuracy of positions of the lens, 
the images, or both, for models with shear, reproducing flux ratios.

\begin{table}
\centering
 \begin{minipage}{84mm}
  \caption{Dependence of the acceptability of fits on required model accuracy}
  \begin{tabular}{@{}ccccc@{}}
  \hline
 $\sigma_\mathrm{I}$ & 0.01 & 0.01  & 0.0025 & 0.0025\\
 $\sigma_\mathrm{L}$ & 0.1 & 0.025  & 0.1 & 0.025\\
 \hline
 {\it Case 1} & $0.43\pm0.17$ & $0.39\pm0.18$ & $0.18\pm0.10$ & $0.16\pm0.10$\\
 {\it Case 2} & $0.59\pm0.22$ & $0.47\pm0.18$ & $0.34\pm0.20$ & $0.28\pm0.17$\\
 {\it Case 3} & $.9997\pm .0002$ & $.9997\pm .0002$ & $.9997\pm .0001$ & $.998\pm .001$\\
 \hline
\noalign{\vskip3pt}
\multicolumn{5}{p{8.4cm}}{Note: The table shows the dependence of 
the rate of acceptability of fits on the method of treating the light
propagation ({\it Cases 1 -- 3}, see the text for details) and on the
assumed errors in
measured positions of the center of the lens ($\sigma_\mathrm{L}$) 
and of the images ($\sigma_\mathrm{I}$). All models  
reproduce flux ratios ($\sigma_\mathrm{m}=0.1$)}  
\end{tabular}
\end{minipage}
\label{accept3}
\end{table}


In the following Fig.~\ref{gamma} we show the probability $P(>\gamma)$ that 
an acceptable fit uses external shear value greater than $\gamma$. 
The plots refer to fits taking  into account image positions but not 
their flux ratios.
\begin{figure}
\includegraphics[width=84mm]{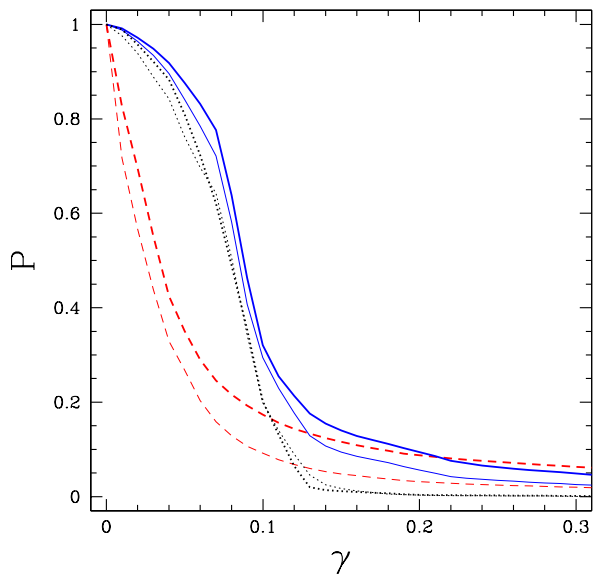}
\caption{The integral probability distribution of 
the external shear value $\gamma$ for acceptable models. 
The results for models neglecting flux ratios ($\sigma_\mathrm{m}=10$)
in {\it Case 1} are shown as solid lines (blue), 
and in {\it Case 2} - as dashed lines (red). For comparison the results 
for models taking into account flux ratios ($\sigma_\mathrm{m}=0.1$)
in {\it Case 1} are shown as dotted lines (black).
The thick lines correspond to lower accuracy ($\sigma_\mathrm{I}=0.01$, 
$\sigma_\mathrm{L}=0.1$) and the thin lines - to models of higher accuracy 
($\sigma_\mathrm{I}=0.0025$, $\sigma_\mathrm{L}=0.025$).}
\label{gamma}
\end{figure}

The influence of lens neighbours and matter in the lens foreground/background
is also illustrated by the changes in the intrinsic lens parameters when 
comparing the fitted model with the original lens used in the simulation.
In reality one has only one set of lens parameters obtained from modeling
and nothing to compare with, nevertheless our plots show the likely 
systematic errors resulting from the model approximations.

\begin{figure}
\includegraphics[width=84mm]{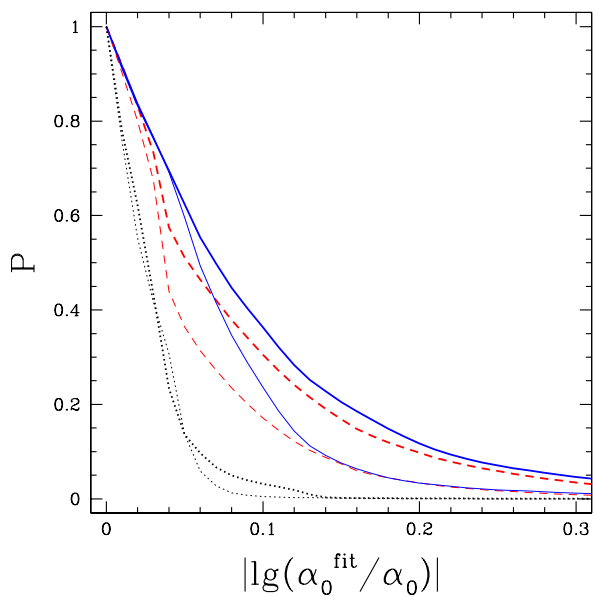}
\caption{The integral probability distribution showing the errors in the 
fitted values of characteristics deflection angle $\alpha_0$.
Conventions follow Fig.~\ref{gamma}.
$\alpha_0$ is the deflection angle of the lens used in simulations.
$\alpha_0^\mathrm{fit}$ is its fitted value.}
\label{alpha}
\end{figure}

\begin{figure}
\includegraphics[width=84mm]{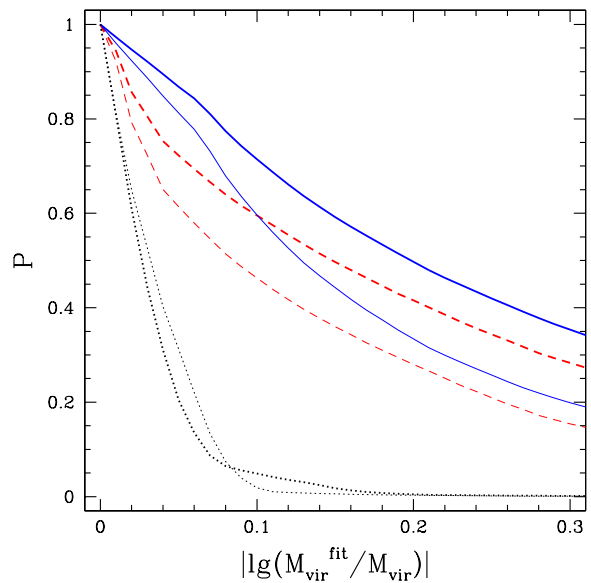}
\caption{The integral probability distribution showing the errors in the 
fitted values of the lens virial mass. Line/colour conventions follow 
Fig.~\ref{gamma}.
$M_\mathrm{vir}$ is the virial mass of the lens used in simulations.
$M_\mathrm{vir}^\mathrm{fit}$ is its fitted value.}
\label{mass}
\end{figure}

In Fig.~\ref{alpha} we show (on a logarithmic graph) the integral 
distribution of the ratios of the fitted values of the deflection angle 
parameter to its original values used for the same lenses in simulations.
 Similarly Fig.~\ref{mass} shows the distribution of the ratios between 
fitted and original virial masses.

\section{Discussion and conclusions}

In simulations we replace each halo taken from the Millennium Simulation 
by a specific kind of matter distribution with the same mass and size,
acting as non-singular isothermal ellipsoid with finite mass. We believe 
that this particular  choice of lens model has little impact on our 
results, because our aim  is to investigate the influence of the 
environment on the lens action, not the action itself. In particular
a different choice of the relative size of the lens core (we always 
use $r_1/r_2=0.01$) may change the properties of the radial caustic
of each lens, but has no influence on the action of matter outside it.

We consider lenses with non-singular matter distribution so the number 
of images should be odd in all cases. In our calculations (which start
from an approximate solution of lens equation) the solutions with four
images are rare, 40 times less frequent than the five-image cases.
The brightest images in our four-image configurations are on average brighter
by $\approx 1.5^\mathrm{mag}$ as compared to the brightest images in 
five-image cases. Also the flux ratio of the brightest image to the next one
is substantially higher (by $\approx 0.6^\mathrm{mag}$) in the four-image
configurations. This suggests that our four-image solutions correspond
to source positions close to the caustics, which produce close image pairs 
unresolvable by our approximate method of image finding. 
Since four-image configurations are rare, they have little impact 
on the statistics of the acceptability of fits.

The matter belonging to gravitationally bound haloes makes only a part 
of the whole matter distribution. The distribution of the unbound 
component we are using is known with limited resolution of 
$\approx 2~h^{-1}\mathrm{Mpc}$ in 3D. Since the influence of low density
regions averages along the lines of sight on distances of several
thousands megaparsecs, the averaging inside cells of few megaparsecs
across probably has little meaning.

Our simulations cover eight small ($3^\prime\times 3^\prime$), randomly 
chosen regions in the sky. In each of the regions we choose ten lenses, 
with largest Einstein ring for a source at $z \approx 2$ according to 
approximate, {\it SIS} lens formula. These are the strongest lenses in 
the region, giving the largest image separations and having the largest area
closed by caustics in the source plane. That does not mean that every 
lensed source must be related to one of the lenses investigated, 
but for its random position in the region of interest this would happen 
in a vast majority of cases. 

Our numerical experiment shows that matter close to the line of sight and 
matter in the close vicinity of the lens do have an important influence
on its lensing properties. Our statement is of statistical nature: we have 
shown than in large part of image configurations obtained as a result of 
simulations taking into account the presence of matter concentrations near 
the lens it is impossible to obtain acceptable models neglecting their 
influence. The problem is more severe in case when simulations take 
into account  not only lens close neighbours but also matter 
inhomogeneities along the beam. It shows that both close neighbours and 
matter along the line of sight are important. Our results are in general
agreement with \citet{WBO04,WBO05,AN11}, but we address different aspect
of the problem.

The quantitative characteristics of the effect depends on the required 
accuracy of the simplified modeling. It certainly depends on the source 
redshift as well, but we have not pursued this aspect of the problem,
limiting ourselves to sources at $z_\mathrm{s}\approx2$.

Some examples of the expected rate of success, when using simplified lens
model to fit realistic image configurations, are shown in 
Tables.~1 and ~2. We assume that every position of the source has the 
same probability. The selection of a sample of multiple 
image systems may also depend on the lens amplification, image configuration 
and other secondary issues, which are difficult to simulate and are 
neglected.

Our results in Tables.~1 and ~2 take into account the fifth image, 
if present.
In real cases the fifth images are not observed or very faint and difficult to
use in modeling. We have performed some extra calculations neglecting 
the fifth images altogether. Of course it is always easier to model a less 
complicated configuration of images, so the rate of success is  higher 
in this case. For instance the first column of Table.~1 would read:
0.16, 0.48, 0.9996 (instead of 0.08, 0.38, 0.9998) and the third column:
0.64, 0.68, 0.9995 (instead of 0.43, 0.59, 0.9997). The changes are 
substantial, but do not qualitatively affect our conclusion.

Our calculations give typical values of the external shear, which is used 
by the acceptable simplified models. When modeling configurations resulting
from the full approach (taking into account all matter inhomogeneities
along the rays) we obtain median shear value $\gamma\approx0.09$ 
(compare Fig.~\ref{gamma}). For the single plane approach (which neglects 
lens foreground and background objects) we get $\gamma\approx0.03$.  
Both numbers weakly depend on the required accuracy of modeling  
and do not change when the fifth image is neglected.
Comparison of these two numbers and the  inspection of Table.~1 suggest 
that shear produced by the matter along  the beam and the matter in the 
lens vicinity cannot be neglected in majority  of cases. It also shows 
that contribution to the shear by background and  foreground objects 
is significant.

We have also compared the fitted values of lens parameters with the 
parameters of the lens used in the simulation. If the flux 
ratios are not modeled, the fitted parameters may substantially differ
from their original values. For instance the characteristic bending 
angle of the  lens $\alpha_0$ may differ by a factor greater than 
$\approx 1.2$ in 50\%  of cases (compare Fig.~\ref{alpha}). 
This would imply similar errors  in estimates of mass distribution 
in the central part of the lens.  For the virial mass 
($M_\mathrm{vir}\propto\alpha_0 r_\mathrm{vir}$) 
the discrepancies may exceed factor $\approx1.5$ in 50\% of cases
(compare Fig.~\ref{mass}). The models reproducing also the flux ratios 
give much better accuracy of the intrinsic lens parameters fitting
(dotted lines in Fig.~\ref{alpha},~\ref{mass}).

\section*{Acknowledgments}
We are grateful to the Anonymous Referee, whose critical remarks greatly 
improved the paper.
The Millennium Simulation databases used in this paper
and the web application providing on-line access to them were constructed
as part of the activities of the German Astrophysical Virtual Observatory.
We are grateful to Volker Springel for providing us with the smoothed
Millennium density distribution in the early stage of this project. 
This work has been supported in part by the Polish 
National Science Centre grant N N203 581540.

\label{lastpage}

\end{document}